\documentclass[longauth,a4paper]{aa}

\usepackage{graphicx}
\usepackage{txfonts}

\newcommand{\etal}{\hbox{et al.} }

\begin{document}

\title{\bf Constraints on the multi-TeV particle population in the Coma Galaxy Cluster with H.E.S.S. observations}

\titlerunning{TeV observations of the Coma galaxy cluster}
\authorrunning{Aharonian \etal}

\date{Received / Accepted}

\offprints{growell@physics.adelaide.edu.au}
\authorrunning{Aharonian \etal}

\author{F. Aharonian\inst{1,13}
 \and A.G.~Akhperjanian \inst{2}
 \and G.~Anton \inst{16}
 \and U.~Barres de Almeida \inst{8} \thanks{supported by CAPES Foundation, Ministry of Education of Brazil}
 \and A.R.~Bazer-Bachi \inst{3}
 \and Y.~Becherini \inst{12}
 \and B.~Behera \inst{14}
 \and K.~Bernl\"ohr \inst{1,5}
 \and C.~Boisson \inst{6}
 \and A.~Bochow \inst{1}
% ...
}

\institute{
Max-Planck-Institut f\"ur Kernphysik, P.O. Box 103980, D 69029
Heidelberg, Germany
\and
 Yerevan Physics Institute, 2 Alikhanian Brothers St., 375036 Yerevan,
Armenia
\and
Centre d'Etude Spatiale des Rayonnements, CNRS/UPS, 9 av. du Colonel Roche, BP
4346, F-31029 Toulouse Cedex 4, France
\and
Universit\"at Hamburg, Institut f\"ur Experimentalphysik, Luruper Chaussee
149, D 22761 Hamburg, Germany
\and
Institut f\"ur Physik, Humboldt-Universit\"at zu Berlin, Newtonstr. 15,
D 12489 Berlin, Germany
\and
LUTH, Observatoire de Paris, CNRS, Universit\'e Paris Diderot, 5 Place Jules Janssen, 92190 Meudon, 
France
\and
IRFU/DSM/CEA, CE Saclay, F-91191
Gif-sur-Yvette, Cedex, France
\and
University of Durham, Department of Physics, South Road, Durham DH1 3LE,
U.K.
\and
Unit for Space Physics, North-West University, Potchefstroom 2520,
    South Africa
\and
Laboratoire Leprince-Ringuet, Ecole Polytechnique, CNRS/IN2P3,
 F-91128 Palaiseau, France
\and 
Laboratoire d'Annecy-le-Vieux de Physique des Particules, CNRS/IN2P3,
9 Chemin de Bellevue - BP 110 F-74941 Annecy-le-Vieux Cedex, France
\and
Astroparticule et Cosmologie (APC), CNRS, Universite Paris 7 Denis Diderot,
10, rue Alice Domon et Leonie Duquet, F-75205 Paris Cedex 13, France
\thanks{UMR 7164 (CNRS, Universit\'e Paris VII, CEA, Observatoire de Paris)}
\and
Dublin Institute for Advanced Studies, 5 Merrion Square, Dublin 2,
Ireland
\and
Landessternwarte, Universit\"at Heidelberg, K\"onigstuhl, D 69117 Heidelberg, Germany
\and
Laboratoire de Physique Th\'eorique et Astroparticules, 
Universit\'e Montpellier 2, CNRS/IN2P3, CC 70, Place Eug\`ene Bataillon, F-34095
Montpellier Cedex 5, France
\and
Universit\"at Erlangen-N\"urnberg, Physikalisches Institut, Erwin-Rommel-Str. 1,
D 91058 Erlangen, Germany
% ...
}

\date{Received / Accepted}


\abstract
{}
{Galaxy clusters are key targets in the search for ultra high energy particle accelerators. The Coma cluster represents one of the
  best candidates for such a search owing to its high mass, proximity, and the established non-thermal radio  
  emission centred on the cluster core.}
{The H.E.S.S. (High Energy Stereoscopic System) telescopes observed Coma for $\sim$8~hr in a search
 for $\gamma$-ray emission at energies $>1$~TeV. The large 3.5$^\circ$ FWHM field of view of H.E.S.S. is ideal for
 viewing a range of targets at various sizes including the Coma cluster core, the radio-relic (1253+275) and merger/infall (NGC~4839)
 regions to the southwest, and features greater than $1^\circ$ away.} 
{No evidence for point-like nor extended TeV $\gamma$-ray emission was found and upper limits to the TeV flux $F(E)$
  for $E>1$, $>5$, and $>10$~TeV were set for the Coma core and other regions. Converting these limits to an energy flux $E^2F(E)$
  the lowest or most constraining is the
  $E>5$~TeV upper limit for the Coma core (0.2$^\circ$ radius) at $\sim$8\% Crab flux units or $\sim10^{-13}$ph~cm$^{-2}$~s$^{-1}$.}
{The upper limits for the Coma core were compared with a prediction for the $\gamma$-ray emission from proton--proton 
  interactions, the level of which ultimately scales with the mass of the Coma cluster. A direct constraint using our most stringent limit 
  for $E>$5 TeV, on the total energy content in non-thermal protons with injection energy spectrum $\propto E^{-2.1}$ and spatial distribution 
  following the thermal gas in the cluster, is found to be
  $\sim$0.2 times the thermal energy, or $\sim 10^{62}$~erg. The $E>$5 TeV $\gamma$-ray threshold in this case corresponds 
  to cosmic-ray proton energies $\ga$50~TeV. Our upper limits rule out the most optimistic theoretical models for
  gamma ray emission from clusters and complement radio observations which constrain the cosmic ray content in clusters 
  at significantly lower proton energies, subject to assumptions on the magnetic field strength.}
{}
\maketitle

\section{Introduction}

Clusters of galaxies represent the largest gravitationally bound objects in the Universe
and are thought to be ideal sites for the acceleration
of particles. The very long confinement time (of order the Hubble time) of the accelerated particles 
(see e.g.  V\"olk \etal \cite{Voelk:1}, Berezinsky \etal \cite{Berezinsky:1}) would allow
interactions of the particles with ambient matter and radiation fields to produce non-thermal
emission from radio to TeV $\gamma$-ray energies. Particles are thought to be accelerated at large-scale 
shocks associated with accretion and merger processes (see e.g. Colafrancesco \etal \cite{Cola:1}, 
Ryu \etal \cite{Ryu:1}), in supernova remnants and galactic-scale winds (V\"olk \etal \cite{Voelk:1}), turbulent 
re-acceleration (Brunetti \& Blasi \cite{Brunetti:2}) and dark matter annihilation (e.g. Colafrancesco \etal \cite{Cola:2}).
In addition, particles may be re-distributed/injected throughout the cluster volume via AGN cluster members 
(En\ss lin \etal \cite{Ensslin:1}, Aharonian \cite{Ahar:1}, Hinton \etal \cite{Hinton:2}),
The non-thermal radio emission observed in recent years from several galaxy clusters (Giovannini \etal \cite{Giovannini:1}, 
Feretti \etal \cite{Feretti:1}) represents clear evidence for relativistic particle populations in such objects. 
Further evidence is provided by possible non-thermal X-rays observed from a few clusters 
(Rephaeli \& Gruber \cite{Rephaeli:1}, Fusco-Femiano \etal \cite{Fusco:1}, Eckert \etal \cite{Eckert:1}).

Gamma-ray emission in galaxy clusters may come from several processes (see review by Blasi \etal \cite{Blasi:2}). 
The collision of relativistic cosmic-ray (CR) protons with thermal nuclei comprising the intra-cluster medium (ICM) may lead to 
$\gamma$-ray emission via the decay of neutral pions (Dennison \cite{Dennison:1}). 
In this context the fraction $\eta$ of thermal energy in the cluster volume in the form of relativistic non-thermal 
particles is an important parameter that can determine the level of $\gamma$-ray emission expected. 
Since the thermal energy content is a function of the cluster mass, the most massive and nearby clusters present the best
opportunity to probe for such $\gamma$-ray emission.
Ultra-relativistic electrons can also up-scatter target photons 
such as the cosmic microwave background (CMB), infrared, starlight, and other soft photon fields)
to TeV $\gamma$-ray energies (Atoyan \& V\"olk \cite{Atoyan:1}, Gabici \etal
\cite{Gabici:1,Gabici:2}). Given that galaxy clusters may accelerate particles to ultra high energies (UHE) 
$> 10^{18}$~eV (e.g. Hillas \cite{Hillas:1}, Kang \etal \cite{Kang:1}), $\gamma$-ray production from inverse-Compton scattering by 
secondary electrons generated when a UHE proton interacts with a CMB photon in the Bethe-Heitler process ($p\gamma \rightarrow e^+e^- + p^\prime$)
may also result (Inoue \etal \cite{Inoue:1}, Kelner \& Aharonian \cite{Kelner:2}). 
Dark matter annihilation has also been considered as a $\gamma$-ray production channel 
(e.g. neutralino annihilation by Colafrancesco \etal \cite{Cola:2}).

Earlier observations in the MeV to GeV $\gamma$-ray band with EGRET only found upper limits only for several clusters 
(Reimer \etal \cite{Reimer:0}) including the Coma cluster. At TeV energies, upper limits 
(Perkins \etal \cite{Perkins:0}) have been reported for the Perseus and Abell~2029 clusters using the single-dish 
{\em Whipple} telescope. The most recent
TeV observations with stereoscopic instruments such as H.E.S.S. (Abell~496 and Abell~85 --- Aharonian \etal \cite{HESS-Abells}), 
and with VERITAS (Coma --- Perkins \etal \cite{Perkins:1}), revealed also upper limits. This work focuses on H.E.S.S.
observations of the Coma cluster. 

Coma (\object{ACO\,1656}) is one of the nearest (z=0.023) and best-studied galaxy clusters.
Extended (several arcminutes in scale) hard X-ray emission (with so far weak evidence for a non-thermal component)
has been observed (Rephaeli \& Gruber \cite{Rephaeli:1}, Fusco-Feminano \etal \cite{Fusco:1}, Rossetti \etal \cite{Rossetti:1},
Lutovinov \etal \cite{Lutovinov:1}, Ajello \etal \cite{Ajello:1}), as well as a prominent non-thermal 
radio halo (Giovannini \etal \cite{Giovannini:1}, Thierbach \etal \cite{Thierbach:1}). The latter is clear evidence for particle
acceleration. 
Being one of the most massive ($M \sim 10^{15} M_{\odot}$) and nearby clusters, with detailed multiwavelength 
observations ranging from low frequency radio wavelength to $\gamma$-rays, the Coma cluster has always been considered as the prototypical
cluster also for very high energy $\gamma$-ray studies. 
The Coma cluster is located in the northern hemisphere and is visible by
H.E.S.S. at moderately high zenith angles (average value $\sim 50^{\circ}$), which leads to a relatively high energy threshold (defined
as the peak detection rate for an $E^{-2.1}$ power-law spectrum of $\gamma$-rays) of
$\gtrsim 1$ TeV. Since the $\gamma$-ray spectrum from clusters is expected to be hard and
extend beyond 10~TeV (basically limited only by the absorption of $\gamma$-ray photons in the cosmic infrared background. At the Coma cluster
distance, an optical depth of unity is reached for energies $E\sim$10 to 20~TeV.), the
energy threshold does not constitute a serious problem for our investigation.

\section{H.E.S.S. observations and analysis}
\label{sec:data}

Operating in the Southern Hemisphere, H.E.S.S. consists of four 
identical 13~m diameter Cherenkov telescopes (Bernl\"ohr \etal \cite{Bernlohr:1}). 
H.E.S.S. employs the stereoscopic imaging atmospheric Cherenkov technique, and is sensitive to $\gamma$-rays 
above an energy threshold of $\sim$0.1~TeV (Hinton \etal \cite{Hinton:1}) for observations at zenith. An angular resolution
of 5 to 6$^\prime$ on an event-by-event basis is achieved, and the large field of view (FoV) with $\mathrm{FWHM}\sim 3.5^\circ$ 
(Aharonian \etal \cite{HESS_Crab}) permits survey coverage in a single pointing. A point source sensitivity of $\sim$1\%~Crab flux 
($\sim10^{-13}$~erg~cm$^{-2}$s$^{-1}$ at 1~TeV) is achieved for a 5$\sigma$ detection after $\sim$25~hr observation. 
Further details concerning H.E.S.S. can be found in Hinton (\cite{Hinton:1}) and references therein.

H.E.S.S. observed Coma during the 2006 season for a total of 8.2~hr (corrected for the detector deadtime) comprising 19 runs
of duration $\sim$28~min each. Those runs were accepted for data analysis if they met the quality control criteria described in 
Aharonian \etal {\cite{HESS_Calibration}.
Data were analysed using the moment-based Hillas analysis procedure described in Aharonian \etal (\cite{HESS_Crab}).
Minimum cuts on the Cherenkov image size\footnote{Total photoelectron signal in the Cherenkov image.} of 80 and 200 photoelectrons corresponding to standard and hard
cuts were employed.
The average zenith angle of the dataset was $\sim$53$^\circ$ yielding energy thresholds (peak detection rate for a power law source
spectrum with an exponent of $2.1$) of $\sim$1.1~TeV and $\sim$2.3~TeV for standard and hard cuts analyses.
This analysis follows on from preliminary H.E.S.S. results (Domainko \etal \cite{Domainko:1}).

The large FoV of H.E.S.S. is well-suited to Coma as TeV emission could be expected from a variety of sites ---
the central radio halo or core; the radio-relic and adjacent galaxy merger/infall region; 
the degree-scale accretion shock suspected to surround the cluster (e.g. Voit \cite{Voit:1}), and individual member galaxies. 
TeV $\gamma$-ray significance skymaps covering a 7$^\circ \times 7^\circ$ FoV (from a mosaic of pointings) 
are presented in Fig.~\ref{fig:tevskymap}), 
employing oversampling radii of 0.2$^\circ$, appropriate for moderately extended sources in the Coma field. Skymaps employing a
0.1$^\circ$ oversampling radius (Fig.~\ref{fig:tevskymap2}) for pointlike sources 
are available in the appendix.
The CR background estimate in skymaps shown here is based on the {\em template}-model 
(Rowell \cite{Rowell:1}), employing a region spatially overlapping the source region but not containing any $\gamma$-ray-like
events. Also available in the appendix are the distributions of skymap significances (Fig.~\ref{fig:1dmaps}) which are well-explained by Gaussians with standard deviation 
within a few percent of unity and means very close to zero, indicating that the background estimate performs well over the
FoV. Similar results were also obtained when employing alternative CR background estimates such as the ring/ring-segment 
and reflected region models (Berge \etal \cite{Berge:1}) which were used for upper limit calculations (in Table~\ref{tab:numbers}).
Results were also cross-checked using an alternative analysis chain. 
\begin{figure*}[ht]
  \centering
    \includegraphics[width=0.49\textwidth]{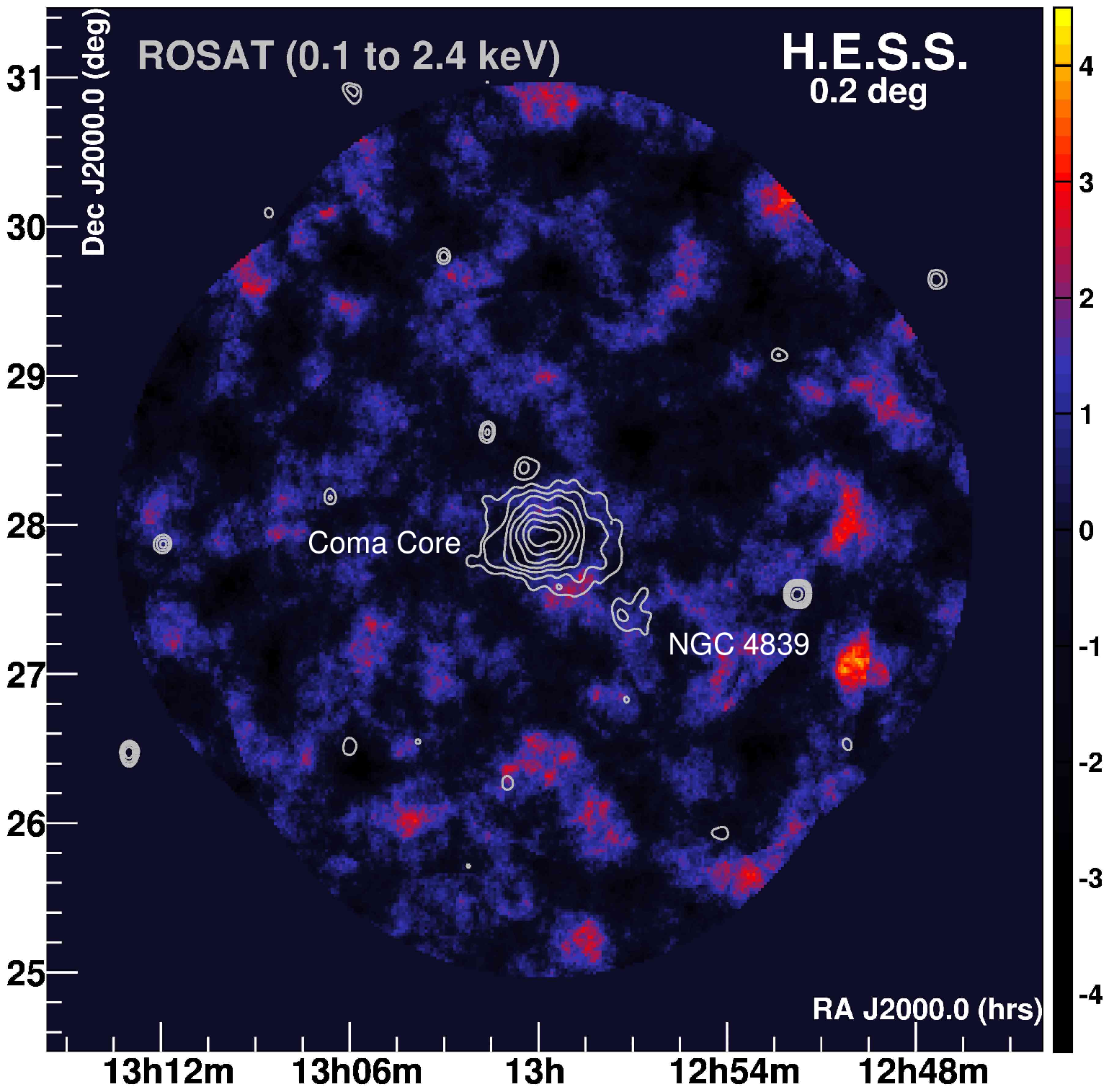} 
    \includegraphics[width=0.49\textwidth]{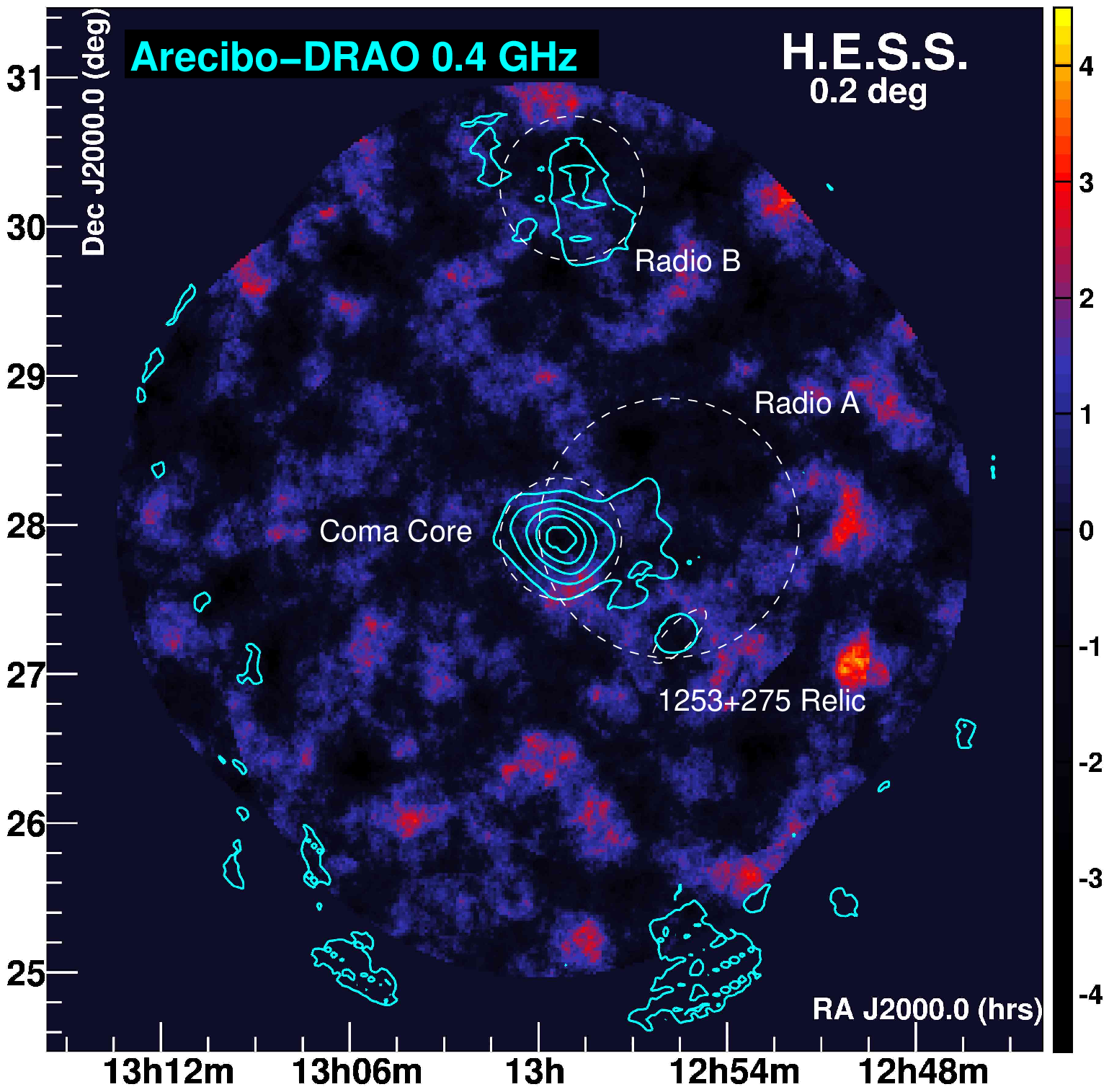} 
    
    \caption{{\bf Left:} Skymap of H.E.S.S TeV excess significance (colour-scale over $\pm4 \sigma$) calculated using Li \& Ma (\cite{Li:1}) 
      over a 7$^\circ \times 7^\circ$
      FoV, employing the template CR background model (Rowell \cite{Rowell:1}). An oversampling radius of 0.2$^\circ$ was used,
      appropriate for extended source searching.  
      Overlaid contours (light-grey solid lines) represent total band 
      (0.1 to 2.4~keV) smoothed X-ray counts~s$^{-1}$ in log-scale from the ROSAT all-sky survey (Voges \etal \cite{Voges:1}).
      {\bf Right:} As for Left but with overlaid contours from radio observations (0.4~GHz - K contours 
      rebinned from the original above 2.9K or $\sim$11$\sigma$) from Kronberg \etal (\cite{Kronberg:1}) 
      with strong point sources removed. The white dashed circle
      indicates the intrinsic 0.4$^\circ$ radius source size and position for the Coma Core (Tab.~\ref{tab:numbers}.)} 
    \label{fig:tevskymap}
\end{figure*}

Table~\ref{tab:numbers} summarises results for various locations in the Coma field guided by results from the ROSAT all sky
survey (Briel \etal \cite{Briel:1}, Voges \etal \cite{Voges:1}), XMM-Newton observations (Feretti \& Neumann \cite{Feretti:2})
and Arecibo-DRAO radio observations (Kronberg \etal \cite{Kronberg:1}).  The H.E.S.S. TeV excess significance $S$ and flux upper limits $\Phi^{99\%}$ (for an $E^{-2.1}$ 
spectrum and using the method of Feldman \& Cousins (\cite{Feldman:1})) for $E>1$, $>5$, and $>10$~TeV were taken from standard, hard, 
and hard cuts analyses respectively. CR background estimates were taken from the reflected model (Berge \etal (\cite{Berge:1})).

In X-rays, extended emission from the Coma cluster and emission further to the southwest are evident. The southwest thermal 
X-ray emission is not entirely spatially coincident with the radio-relic (discussed below), but is centred on the galaxy 
sub-group NGC~4839 (labeled NGC~4839 in Table~\ref{tab:numbers}), $\sim20^\prime$ closer to the Coma cluster core compared to the 
radio-relic. This sub-group is thought to represent a merger or infall of galaxies associated with Coma.
Hard X-ray (18--30~keV) observations with INTEGRAL (Eckert \etal \cite{Eckert:1}) suggest excess emission in the direction 
of this infalling sub-group close to the cluster centre.
In radio, the Coma~core is visible as well as a radio extension to the southwest known as the radio relic region
(labeled 1253+275-Relic in Table~\ref{tab:numbers}). Two additional regions were also chosen to overlap the diffuse radio features 
labeled 'A' and 'B' from Kronberg \etal (\cite{Kronberg:1}) (at 0.4~GHz)
for which circular regions of 
radii 0.9$^\circ$ and 0.5$^\circ$ respectively were used. Radio~'A' appears to well encompass the radio core of Coma which is discussed
at length in Thierbach \etal (\cite{Thierbach:1}) from their $>2$GHz observations, whilst radio~'B' is a new feature from 
Kronberg \etal (\cite{Kronberg:1}).
In all cases no evidence for TeV emission was seen and 99\% confidence level flux upper limits 
(assuming an $E^{-2.1}$ spectrum) at several energy thresholds ($E>1$, $>5$ and $>10$~TeV) were set. We note the highest excess significance feature 
  at $\sim 4.1\sigma$ towards RA=12$^h$55$^m$ Dec=$+$27$^\circ$15$^\prime$ is expected by chance given the number of independent trials ($\sim 10^5$) in the image.
\begin{table*}
  \caption{Numerical summary for various regions in the Coma galaxy cluster and surrounding field.}
  \centering
  {\small
  \begin{tabular}{llccccrrrrrr}
   Name & 1RXS & R.A. & Dec & $^1$RoI [deg] & $^2T$ [h] & \multicolumn{3}{c}{$^3$$S$ [$\sigma$]} & \multicolumn{3}{c}
   {Flux U.L. $^4\Phi^{99\%}$} \\  
        &      &  [J2000.0] & [J2000.0]          &               &         & \multicolumn{3}{c}{($E>$1,5,10~TeV)}     & \multicolumn{3}{c}
   {($E>$1,5,10~TeV)} \\[1mm] \hline \\[-2mm]
   Coma~Core        & J125947.9+275636 & 12$^h$59$^m$47.9$^s$  & 27$^\circ$56$^\prime$36$^{\prime\prime}$ & 0.0           & 7.3    & -0.5 & -1.2 & -1.4 &  6.1 & 0.3 & 0.1 \\ 
                    &                  &           &          & 0.2           & 7.3    & +0.4 & -0.4 & -0.6 & 10.8 & 0.9 & 0.5\\
                    &                  &           &          & 0.4           & 7.3    & +1.1 & -0.5 & -1.5 & 25.5 & 1.7 & 0.6\\
   1253+275-Relic$^\S$ &               & 12$^h$55$^m$15.0$^s$ & 27$^\circ$15$^\prime$00$^{\prime\prime}$    & $\S$          & 5.7    & +1.3 & +2.0 & +1.0   & 15.9 & 2.6 & 1.2\\ 
   Radio-A          &                  & 12$^h$55$^m$00.0$^s$ & 28$^\circ$00$^\prime$00$^{\prime\prime}$   & 0.9           & 4.6    & +2.0 & -2.2 & -2.0 & 78.7 & 2.3 & 1.4\\ 
   Radio-B          &                  & 13$^h$00$^m$00.0$^s$ & 30$^\circ$15$^\prime$00$^{\prime\prime}$    & 0.5           & 1.6    & +2.5 & +0.7 & +0.0 & 77.8 & 7.0 & 3.4\\ 
   NGC~4839         & J125710.8+272426 & 12$^h$57$^m$24.3$^s$ & 27$^\circ$29$^\prime$52$^{\prime\prime}$ & 0.2           & 6.8    & +0.4 & +1.7 & +1.2 &  9.0 & 1.9 & 1.0\\ 
                    &                  &           &          & 0.4           & 6.8    & -1.8 & +0.3 & +0.1 &  6.7 & 2.0 & 1.3\\[1mm] \hline
   \multicolumn{12}{l}{\scriptsize  $1.$ Source Region of Interest (RoI) intrinsic radius. Actual radii used are convolved with the 
     analysis PSF. A zero value here refers to a point-source analysis.}\\
   \multicolumn{12}{l}{\scriptsize  $2.$ Observation time (hr) corrected to a 0.7$^\circ$ off-axis angle using a standard cuts $E>1$~TeV response curve.}\\
   \multicolumn{12}{l}{\scriptsize  $3.$ Statistical significance using Li \& Ma (\cite{Li:1})}\\
   \multicolumn{12}{l}{\scriptsize  $4.$ 99\% C.L. flux upper limit $\times 10^{-13}$~ph cm$^{-2}$ s$^{-1}$} \\
   \multicolumn{12}{l}{\scriptsize  $\S$ Elliptical region ($0.33^\circ \times 0.2^\circ$ with position angle 45$^\circ$) 
     as defined in Feretti \& Neumann (\cite{Feretti:2})}\\
  \end{tabular}
  }
  \label{tab:numbers}
\end{table*}

\section{Discussion}
\label{sec:discussion}

One of the most important properties of clusters of galaxies is the fact that CR protons remain diffusively confined in 
the magnetised intracluster medium for cosmological time scales.
The maximum energy that can be confined depends on the (unknown) diffusion coefficient but an often made assumption is that the
maximum energy  
is well above that relevant for TeV $\gamma$-ray emission (V\"olk \etal \cite{Voelk:1}).
CR protons lose their energy mainly via proton--proton interactions in the intergalactic medium. Due to the low density 
of this medium, the energy loss time is longer than the Hubble time. This implies that the hadronic CR content 
of a cluster is simply the superposition of the contributions from all the CR sources which have been active during the 
cluster lifetime, with little attenuation due to energy losses.
Under reasonable assumptions on the CR acceleration efficiency, the total non-thermal energy stored 
in the intracluster medium might be of the same order of magnitude of the thermal energy. For example Ryu \etal (\cite{Ryu:1})  
have estimated a non-thermal energy fraction reaching 50\% of the thermal energy.
Such an amount of CR protons would result in copious emission of $\gamma$-rays from the decay of neutral pions produced 
in proton--proton interactions.
Since the most optimistic theoretical predictions are well within the capabilities of current-generation Cherenkov telescopes, the upper limits 
obtained by H.E.S.S. can be used effectively to constrain the non-thermal energy content of the Coma cluster.

A remarkable feature of the $\gamma$-ray emission from neutral pion decay is that its spatial profile is expected to follow the 
density profile of the gas which constitutes the target for proton--proton interactions.
In the case of the Coma cluster, this gas is concentrated within a core of radius $\sim 300$~kpc which, at the distance of 
the Coma cluster, corresponds to $\sim 0.2^{\circ}$ (see discussion below). This is the basis for the angular regions from which upper limits have 
been extracted.

Table~\ref{tab:limits} demonstrates how the upper limits on the $\gamma$-ray emission convert into upper limits on the 
ratio of the cluster thermal energy to that of CR protons (non-thermal energy). This non-thermal to thermal energy
ratio is denoted $\eta = E_{CR}/E_{th}$.
The Coma cluster thermal energy has been evaluated using the gas density profile and the intracluster medium temperature 
derived from X-ray data (e.g. Neumann \etal \cite{Neumann:1}) and resulted in $E_{th} \sim 3.9 \times 10^{62}$ and 
$1.4 \times 10^{63}$~erg for regions within 0.2 and 0.4 degrees from the cluster centre respectively.
The expected $\gamma$-ray emission has been computed following Kelner \etal (\cite{Kelner:1}) and assuming that the energy spectrum 
for CR protons is a single power law with spectral index $\alpha = 2.1$ and 2.3 starting at an energy of 1~GeV. 
The assumption of such hard spectra is justified by the 
fact that, due to CR confinement within the intracluster medium, the equilibrium spectrum must be equal to the CR injection 
spectrum at the sources. Note that much steeper spectra ($\alpha$ up to 6) are indicated for CRs accelerated at weak merger
shocks (Gabici \& Blasi \cite{Gabici:1b}, Berrington \& Dermer \cite{Berrington:1}).
The cluster non-thermal energy has been obtained by integrating the spectrum above 1 GeV and the resulting 
$\gamma$-ray emission corrected for absorption in the cosmic infrared background (CIB) using the Salpeter initial mass function 
opacity given in Primack (\cite{Primack:1}). More recent constraints on the CIB from the TeV blazar 1ES~0229+200 
(Aharonian \etal \cite{HESS-0229})
provide only a negligible change in absorption from 1 to 10~TeV given Coma's proximity. 
Our upper limits can then be used to constrain this overall CR spectrum and hence non-thermal energy.

The only missing piece of information is the spatial distribution of CRs. This quantity is not unambiguously known and it depends 
on the spatial distribution of CR sources in clusters. Here, two distinct situations were considered 
as two extreme cases. In the first one, referred to as Model-A, the radial profile of the CR energy density was assumed to 
follow the thermal energy profile. In Model-B, a spatially homogeneous distribution for the CRs was assumed.

\begin{table}
  \caption{Constraints$^\dagger$ on the ratio of CR (non-thermal) to thermal energy ($E_{CR}/E_{th}$) 
    for the Coma cluster core region (within two radii) and assumed  
    cosmic-ray distribution models A and B (see text).} 
  \centering
  {\small
  \begin{tabular}{ccccc}
    Radius & $\alpha$ & Model & $\eta = E_{CR}/E_{th}$ & $E_{CR}$ [erg] \\[1mm] \hline \\[-2mm]
       0.2$^\circ$ (0.33 Mpc)  & 2.1           & A             &     $<$0.19   &  $<$7.4$\times 10^{61}$ \\   
       0.4$^\circ$ (0.67 Mpc)  & 2.1           & A             &     $<$0.18   &  $<$2.5$\times 10^{62}$ \\   
       0.4$^\circ$ (0.67 Mpc)  & 2.1           & B             &     $<$0.25   &  $<$3.5$\times 10^{62}$ \\
       0.4$^\circ$ (0.67 Mpc)  & 2.3           & A             &     $<$0.55   &  $<$7.7$\times 10^{62}$ \\[1mm] \hline
       \multicolumn{5}{l}{\scriptsize $\dagger$ The upper limit for $E>5$~TeV has been used here.}\\
     \end{tabular}
  }
  \label{tab:limits}
\end{table}

In all cases considered, the most constraining data points in terms of energy flux are the upper limits for photon energies 
above 5~TeV. Our $>$10~TeV limits are marginally higher ($\sim$20\%) whilst the $>$1~TeV limits are factor 2 to 3 higher. 
For a CR spectrum with $\alpha = 2.1$ and assuming that the CR energy density follows the thermal energy 
density (Model-A) values for $\eta \lesssim 0.2$ for both the considered regions ($0.2^{\circ}$ and $0.4^{\circ}$) were obtained. 
In order to check how the upper limits depend on the assumptions made on the CR spectrum and spatial distribution, the 
size of the region was fixed to $0.4^{\circ}$ and two more cases were considered: a homogeneous distribution of CRs (Model-B) and a softer 
CR spectrum with $\alpha = 2.3$. The upper limits on $\eta$ are in these cases less stringent and are reported in the third 
and fourth rows of Table~\ref{tab:limits}, resulting in $\eta \lesssim 0.25$ and $0.55$ respectively.
In combination with our limits for $E>1$~TeV and $>$10~TeV, model-independent constraints on the 
$\ga 10$, $\ga 50$, $\ga 100$~TeV CR proton population in Coma were set.

Our upper limits can be compared with those obtained from observations at other wavelengths and also other models.
Firstly, our limits are slightly more constraining compared to those obtained from EGRET MeV/GeV data 
($\eta$=0.45 and 0.25 assuming $\alpha$=2.1, 2.3; Pfrommer \& En\ss lin \cite{Pfrommer:1}).
Our limits also rule out some of the models for CR acceleration in clusters of galaxies that predict high $\eta$ values, 
even up to 50\% (Ryu \etal \cite{Ryu:1}). The high frequency non-thermal radio emission of the Coma cluster 
(Thierbach \etal \cite{Thierbach:1}) has been used by Reimer \etal (\cite{Reimer:1}) to constrain $\eta$ by noticing that 
the radio emission from 
secondary electrons produced by CRs in proton--proton interactions cannot exceed the measured value. 
Recent observations (Brunetti \etal \cite{Brunetti:1}) of a steep radio spectrum 
from the cluster Abell~521 have also been considered in a similar way as in the Reimer \etal work on Coma.
Their rather stringent limits obtained using radio data  ($\eta$=10$^{-4}$ to $\sim$0.3) depend quadratically on the value of the 
intracluster $B$-field which has a large uncertainty of a factor $\sim$10 ($B=$0.1 to 2$\mu$G). 
Additionally the $<10$~GHz radio measurement constrains formally the $<0.1$~TeV CR population within the range of $B$-fields used. 
ULs from H.E.S.S. and other VHE gamma-ray instruments make direct constraints on the $E\ga 10$~TeV CR population, energies well above 
that implied by Reimer \etal (\cite{Reimer:1}), and are essentially independent of the $B$-field. Note that an additional preliminary 
$E>0.3$~TeV upper limit from $\sim$19~hr of VERITAS (Perkins \etal \cite{Perkins:1}) observations for the Coma core (0.3$^\circ$ radius) has been reported at
$\sim$3\% Crab (or $\sim 2\times 10^{-12}$~ph~cm$^{-2}$~s$^{-1}$). This VERITAS limit, when converted to an energy flux, provides a constraint on 
$\eta$ very similar to ours, albeit for CRs of slightly lower energies $\ga$few~TeV.
Finally, our $\eta$ constraints for Coma are within a factor of two to three larger than those obtained from somewhat 
deeper H.E.S.S. observations ($\sim$33~hr) on the Abell~85 cluster ($M=7.6\times 10^{14}$ M$_\odot$; $z=$0.055) 
(Aharonian \etal \cite{HESS-Abells}). In this work, the constraints $\eta<0.06$ to $\eta<0.15$ for Abell~85 were reported for a 
range of mass profiles and a CR spectral index of $-$2.1.

\section{Conclusions}
\label{sec:conclusions}

H.E.S.S. observed the Coma galaxy cluster for $\sim$8~hr, obtaining upper limits to the $E>1$, $>5$, and $>10$ TeV 
$\gamma$-ray flux from the cluster core (0.2$^\circ$ and 0.4$^\circ$ radii). Additional regions for which 
upper limits are given include the radio-relic (1253+275); the merger/infall region associated with NGC~4839; and two large-scale 
radio features 'radio-A' (0.9$^\circ$ radius) and 'radio-B' (0.5$^\circ$ radius). 

Our results were compared to a model for the proton-proton $\gamma$-ray emission assuming a proton spatial profile matching 
the centrally peaked thermal gas and injection spectral index of 2.1. In this case our $E>5$~TeV H.E.S.S. upper limit for the 
Coma core region within a 0.2$^\circ$ radius region (amounting to $\sim$8\% Crab flux units or
$\sim10^{-13}$ph~cm$^{-2}$~s$^{-1}$) constrains the fraction of energy $\eta$ in CRs 
to $<$0.2 times the thermal intracluster medium energy, or $\sim$10$^{62}$~erg. This can be compared with
the generally more stringent constraints on $\eta$ so far from radio observations  (Reimer \etal \cite{Reimer:1}) in the range 
$\eta$=10$^{-4}$ to $\sim$0.3 times the thermal energy for $B$-fields 0.1 to 2$\mu$G in the intracluster medium and 
proton injection spectral 
index $\alpha$=2.1 to 2.5. It should be noted that the H.E.S.S. $E>5$~TeV upper limit assuming $\alpha=2.1$ formally constrains $\ga 50$~TeV CRs whilst the
radio limits pertain to $<0.1$~TeV CRs, highlighting the complementarity of the two approaches. 
The H.E.S.S. constraints for the amount of CRs stored in the intracluster medium also rule out the most optimistic theoretical 
models for CR acceleration in clusters of galaxies. 

Our upper limit for the NGC~4839 merger/infall group may also be useful, in conjunction with the hard X-ray emission,
in constraining models for the additional shock acceleration of particles potentially associated with this region. 

Our focus here has been on the expected centrally peaked proton-proton $\gamma$-ray emission, but emission expected from
the inverse-Compton scattering of electrons, and UHE proton/$\gamma$ interactions $p\gamma \rightarrow e^+e^- + p^\prime$ may follow the
spatial profile of the degree-scale cluster accretion shock as an annulus of radius $\sim 1-2^\circ$. Inhomogeneities in the shock
structure leading to TeV hotspots of up to degree in size (Keshet \etal \cite{Keshet:1}), and uncertainties in the shock size 
prevented a specific attempt to search for such emission. 
TeV $\gamma$-ray observations can in principle provide direct constraints on 
the ability of such shocks to accelerate particles to multi-TeV energies and beyond. 
The significance skymap presented in Fig.~\ref{fig:tevskymap} 
for 0.2$^\circ$ radii sources would suggest however that no such evidence for moderate-scale emission in the Coma field is 
present. Indeed, similar skymaps are obtained for a 0.4$^\circ$ radius.

Overall our results indicate that deeper observations (towards $100$ hr) of the Coma cluster, or other similarly massive 
and nearby galaxy cluster in TeV $\gamma$-rays, are warranted to probe the Universe's largest-scale shocks. 
Such a detection appears possible unless the total energy in the form of multi-TeV CRs is significantly less than
$\sim 10$~\% of the cluster thermal energy. 

Finally, the recently launched LAT instrument onboard the Fermi GST will provide critical constraints in the MeV/GeV band
(likely within its first year or so of observation) and these results are eagerly awaited.

\begin{acknowledgements}
  The support of the Namibian authorities and of the University of Namibia
  in facilitating the construction and operation of H.E.S.S. is gratefully
  acknowledged, as is the support by the German Ministry for Education and
  Research (BMBF), the Max Planck Society, the French Ministry for Research,
  the CNRS-IN2P3 and the Astroparticle Interdisciplinary Programme of the
  CNRS, the U.K. Particle Physics and Astronomy Research Council (PPARC),
  the IPNP of the Charles University, the Polish Ministry of Science and
  Higher Education, the South African Department of
  Science and Technology and National Research Foundation, and by the
  University of Namibia. We appreciate the excellent work of the technical
  support staff in Berlin, Durham, Hamburg, Heidelberg, Palaiseau, Paris,
  Saclay, and in Namibia in the construction and operation of the equipment.
  We thank Philipp Kronberg for the 408~MHz radio image.
\end{acknowledgements}

\begin{appendix}

\section{Additional Figures}

H.E.S.S. images as for Fig.\ref{fig:tevskymap} but using a 0.1$^\circ$ integration radius are presented here
as well as distributions of significances for both integration radii.

\begin{figure}[ht!]
  \centering
    \includegraphics[width=0.49\textwidth]{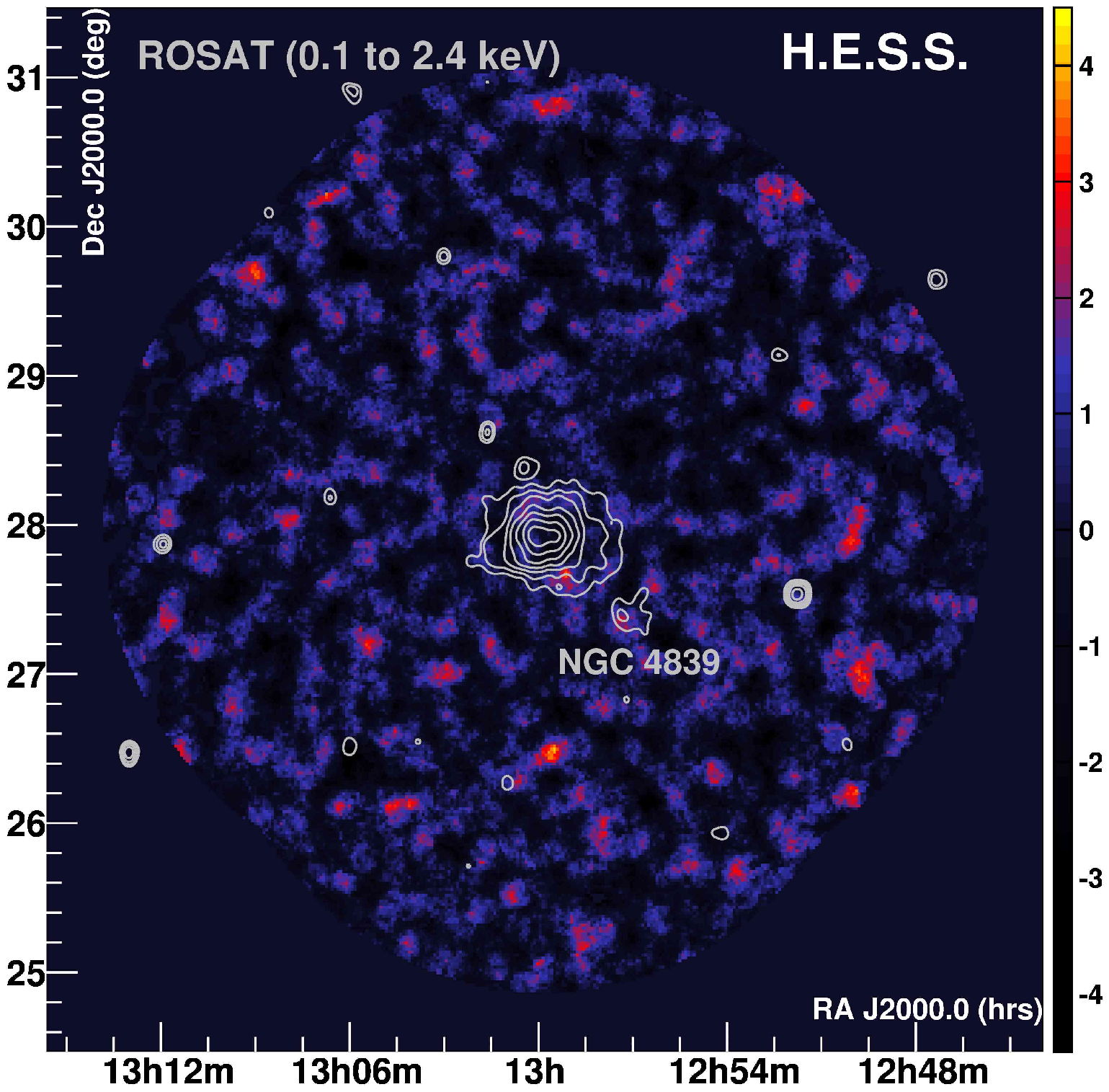}
    \includegraphics[width=0.49\textwidth]{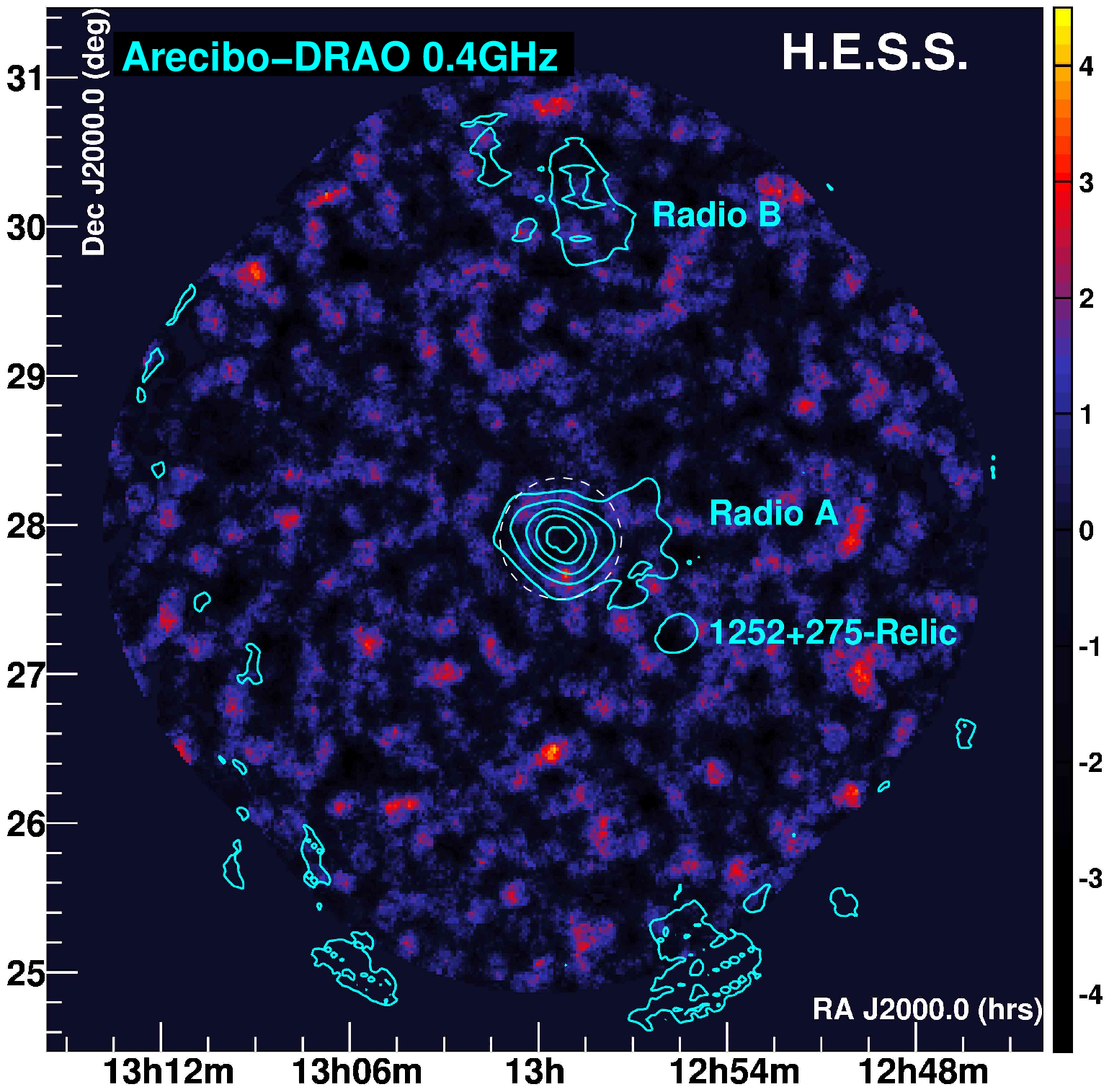}
    
    \caption{H.E.S.S. signficance skymaps as for Fig.\ref{fig:tevskymap} but using an oversampling radius of 0.1$^\circ$. 
      {\bf Top:} Overlaid contours (light-grey solid lines) represent total band 
      (0.1 to 2.4~keV) smoothed X-ray counts~s$^{-1}$ in log-scale from the ROSAT all-sky survey (Voges \etal \cite{Voges:1}).
      {\bf Bottom:} Overlaid contours from radio observations from Kronberg \etal (\cite{Kronberg:1}) with strong point sources removed.}
      \label{fig:tevskymap2}
\end{figure}
\begin{figure}[ht!]
  \centering
  \includegraphics[width=0.45\textwidth]{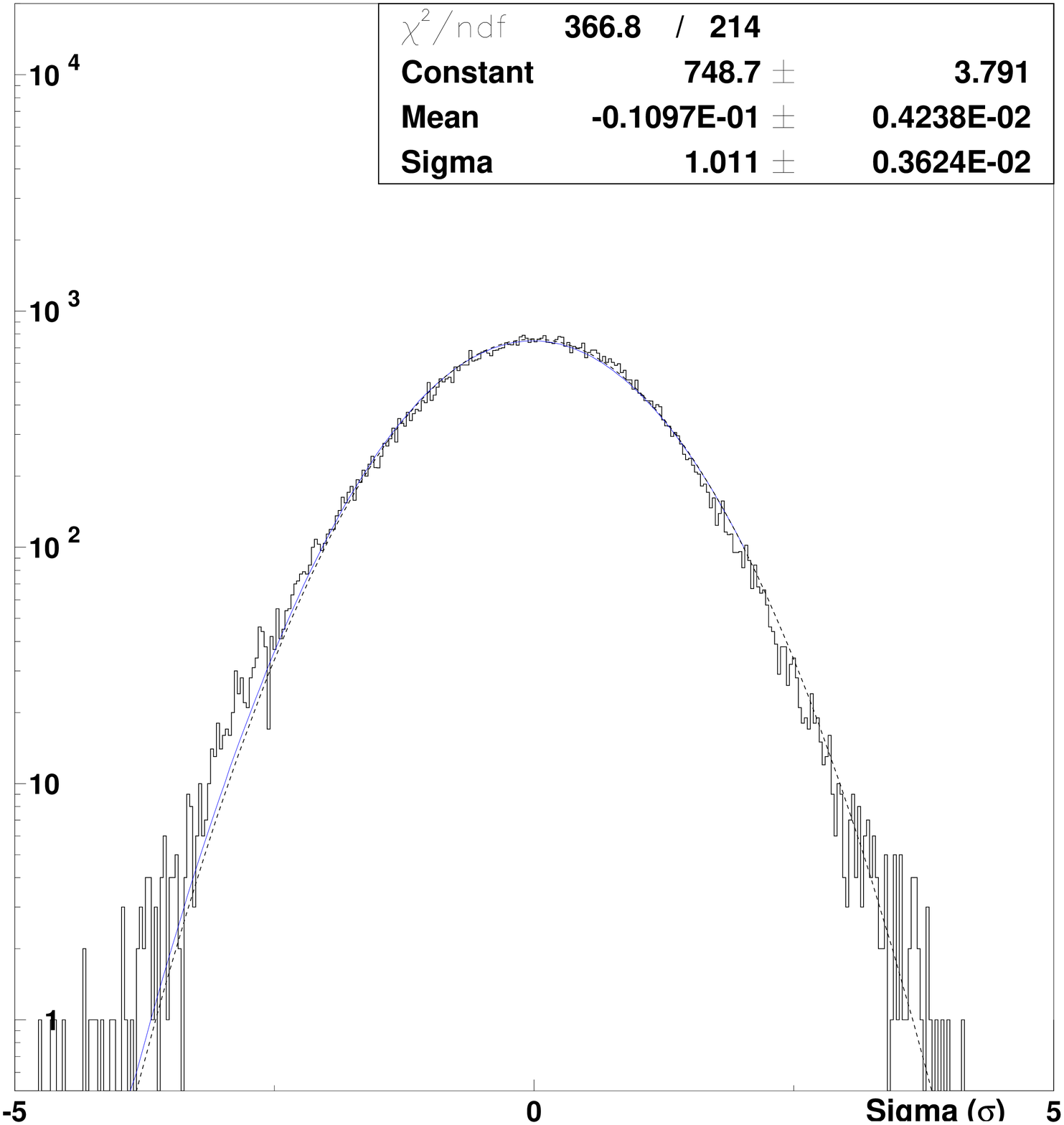}
  \includegraphics[width=0.45\textwidth]{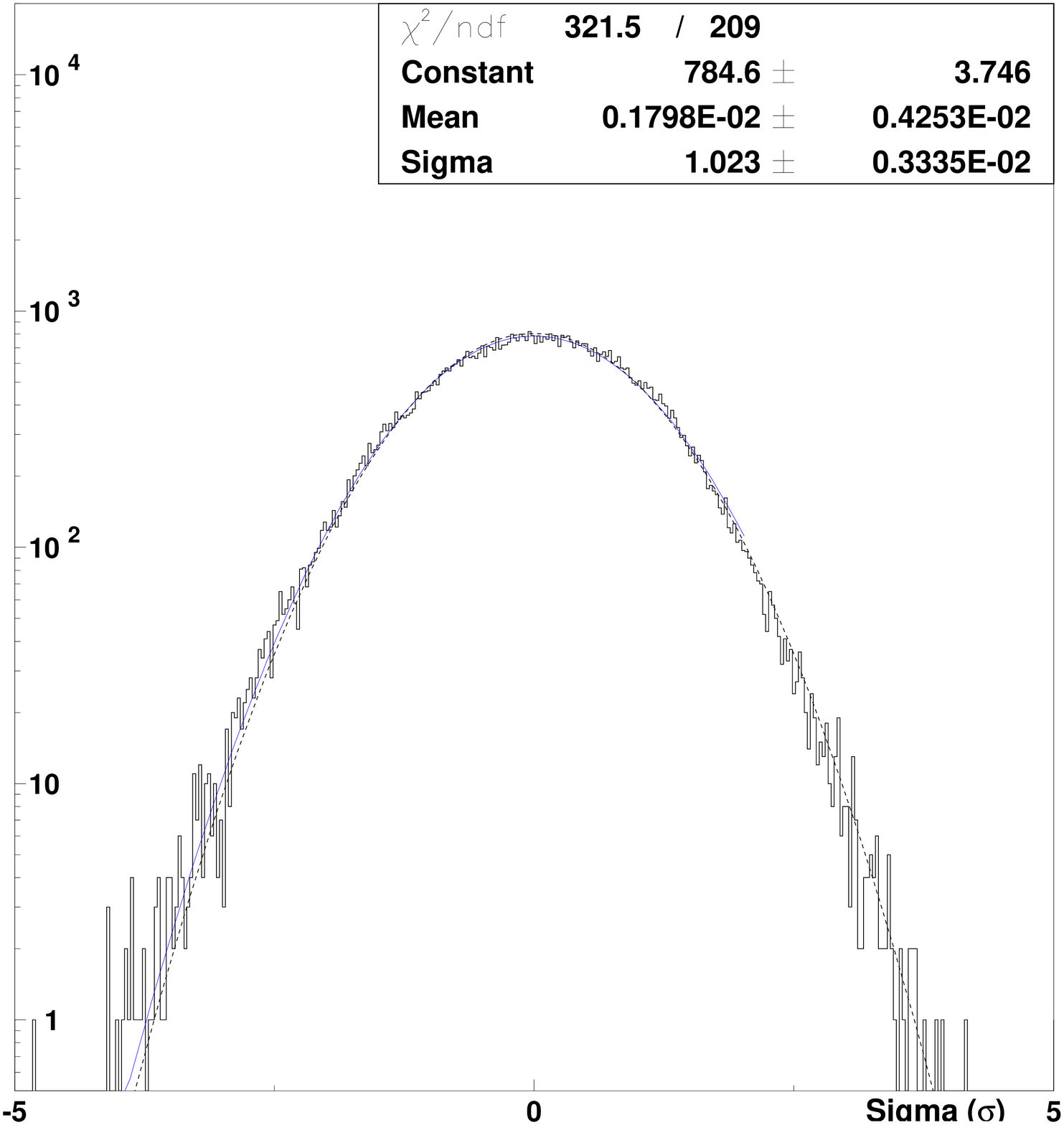}
  \caption{Distributions of H.E.S.S. significance from Figs:~\ref{fig:tevskymap} and \ref{fig:tevskymap2}
    for various oversampling radii --- 
    {\bf Top:} 0.2$^\circ$; {\bf Bottom:} 0.1$^\circ$.The dashed line represents a Gaussian
    of zero mean and unit standard deviation.}
  \label{fig:1dmaps}
\end{figure}

\end{appendix}

\end{document}